\documentclass[twocolumn]{aastex631}
\usepackage{multirow}
\usepackage{graphicx}
\usepackage{txfonts}
\usepackage{natbib}
\usepackage{color}
\usepackage{hyperref}
\hypersetup{colorlinks=True,citecolor=blue,allcolors=blue}
\usepackage{tabularx}
\usepackage{tabu}
\usepackage{float}
\usepackage{academicons}
\usepackage{xcolor}
\usepackage{soul}

\shorttitle{First Direct Evidence for Keplerian Rotation in Quasar Inner BLRs}
\shortauthors{Fian et al.}

\begin{document}
\title{First Direct Evidence for Keplerian Rotation in Quasar Inner Broad Line Regions}

\correspondingauthor{C. Fian}
\email{carina.fian@uv.es}

\author[0000-0002-2306-9372]{C. Fian}
\affiliation{Departamento de Astronom\'{i}a y Astrof\'{i}sica, Universidad de Valencia, E-46100 Burjassot, Valencia, Spain}
\affiliation{Observatorio Astron\'{o}mico, Universidad de Valencia, E-46980 Paterna, Valencia, Spain}

\author[0000-0001-7798-3453]{J. Jim\'enez-Vicente}
\affiliation{Departamento de F\'{\i}sica Te\'orica y del Cosmos, Universidad de Granada, Campus de Fuentenueva, 18071 Granada, Spain}
\affiliation{Instituto Carlos I de F\'{\i}sica Te\'orica y Computacional, Universidad de Granada, 18071 Granada, Spain}

\author[0000-0003-1989-6292]{E. Mediavilla}
\affiliation{Instituto de Astrof\'{\i}sica de Canarias, V\'{\i}a L\'actea S/N, La Laguna 38200, Tenerife, Spain}
\affiliation{Departamento de Astrof\'{\i}sica, Universidad de la Laguna, La Laguna 38200, Tenerife, Spain}

\author[0000-0001-9833-2959]{J. A. Mu\~noz}
\affiliation{Departamento de Astronom\'{i}a y Astrof\'{i}sica, Universidad de Valencia, E-46100 Burjassot, Valencia, Spain}
\affiliation{Observatorio Astron\'{o}mico, Universidad de Valencia, E-46980 Paterna, Valencia, Spain}

\author[0000-0002-4830-7787]{D. Chelouche}
\affiliation{Department of Physics, Faculty of Natural Sciences, University of Haifa, Haifa 3498838, Israel}
\affiliation{Haifa Research Center for Theoretical Physics and Astrophysics, University of Haifa,
Haifa 3498838, Israel}

\author[0000-0002-9925-534X]{S. Kaspi}
\affiliation{School of Physics and Astronomy and Wise Observatory, Raymond and Beverly Sackler Faculty of Exact Sciences, Tel-Aviv University, Tel-Aviv, Israel}

\author[0000-0002-6482-2180]{R. Forés-Toribio}
\affiliation{Departamento de Astronom\'{i}a y Astrof\'{i}sica, Universidad de Valencia, E-46100 Burjassot, Valencia, Spain}
\affiliation{Observatorio Astron\'{o}mico, Universidad de Valencia, E-46980 Paterna, Valencia, Spain}

\begin{abstract}
We introduce a novel method to derive rotation curves with light-day spatial resolution of the inner regions of lensed quasars. We aim to probe the kinematics of the inner part of the broad-line region (BLR) by resolving the microlensing response -- a proxy for the size of the emitting region -- in the wings of the broad emission lines (BELs). Specifically, we assess the strength of the microlensing effects in the wings of the high-ionization lines Si \textsc{IV} and C \textsc{IV} across various velocity bins in five gravitationally lensed quasars: SDSS J1001+5027, SDSS J1004+4112, HE 1104$-$1805, SDSS J1206+4332, and SDSS J1339+1310. Using Bayesian methods to estimate the dimensions of the corresponding emission regions and adopting a Keplerian model as our baseline, we examine the consistency of the hypothesis of disk-like rotation. Our results reveal a monotonic, smooth increase in microlensing magnification with velocity. The deduced velocity-size relationships inferred for the various quasars and emission lines closely conform to the Keplerian model of an inclined disk. This study provides the first direct evidence of Keplerian rotation in the innermost region of quasars across a range of radial distances spanning from $\sim$5 to 20 light-days. 
\end{abstract} 

\section{Introduction} \label{intro}
Direct evidence concerning the kinematics within the regions that emit the characteristic broad emission lines (BELs) in active galactic nuclei (AGNs) and quasars remains scarce. Generally, the shape of these BELs cannot be conclusively attributed to specific kinematic processes. Recent studies (see, e.g., \citealt{Nagoshi2024,Fian2023blr}) support a two-component model for describing the kinematics of the broad line region (BLR), as proposed by \citet{Popovic2004}. This model suggests that the line wings originate from the accretion disk, while the line core emerges from a broadly spherical emission region. However, this framework often requires refinement to capture the complex morphology of the line profiles accurately. In most AGNs and quasars, distinguishing and separating these kinematic components proves challenging. Kinematic analyses typically rely on the virial theorem, which relates the widths of the emission lines with the dimensions of the emitting regions -- usually estimated through a size-luminosity relationship (see, e.g., \citealt{Bentz2009,Kaspi2021}). Nonetheless, such studies primarily focus on the core kinematics, offering limited insight into the innermost region of the BLR.

The commonly observed strong BELs are complex, with contributions from the two aforementioned kinematic components. Exceptions may include the wings of the line profiles, where contributions mainly from the highest projected velocities -- presumably arising from the innermost regions of the BLR -- are expected. A way to experimentally explore the inner BLR kinematics involves establishing a relationship between the high-velocity bins of observed line profiles and the radial distances at which the corresponding high-velocity emitters are located. One way to achieve this is through velocity-resolved time delay measurements using reverberation mapping (see, e.g., \citealt{DeRosa2018,Bentz2023}). This method has been applied primarily to Seyfert galaxies, but results are complex and difficult to interpret. Another possibility is to use spectro-astrometry to spatially separate the photo-centers of the regions corresponding to different wavelength bins of the emission line. This technique has been applied with Gravity@VLTI to 3C273 (\citealt{Gravity2018}), achieving spatial resolution on the order of tens of light-days. The separations between the photo-centers suggest a disk structure, although additional information is needed to unambiguously distinguish between rotational motion and inflow or outflow. As an alternative for measuring the distance at which the emitters of a velocity bin lie with light-day spatial resolution, we propose using gravitational microlensing. This method is sensitive to the size of the emitting region and can extend the study of the BLR kinematics to quasars at intermediate redshifts. As we will demonstrate, it is possible to obtain smooth and well-sampled experimental kinematic curves that relate microlensing-based sizes to velocity bins.

In this study, we examine the impact of microlensing on resolved velocity bins in the wings of the high-ionization lines Si IV and C IV in five lensed systems: SDSS J1001+5027, SDSS J1004+4112, HE 1104$-$1805, SDSS J1206+4332, and SDSS J1339+1339. 
In recent literature, there has been a growing argument that the high-ionization line C IV may arise, at least partially, from a disk-like structure (\citealt{Fian2018blr, Fian2021blr, Fian2023_1004, Hutsemekers2021, Hutsemekers2023,Hutsemekers2024}). Previous works based on microlensing have primarily focused on the emission lines as a whole, without resolving the emission line profile into velocity bins. The objects selected for the present study display significant microlensing-induced distortions in their line profiles, which provide a unique opportunity to investigate the spectral response to microlensing across different kinematic regions within the BLR. As we will demonstrate, we can obtain smooth and well-sampled experimental kinematic curves that link microlensing-based sizes with velocity bins. By comparing the predicted microlensing response of our kinematic model with the observed responses, we aim to probe and constrain the emitter's kinematics and evaluate the hypothesis of Keplerian rotation.

The paper is structured as follows. Sections \ref{datades} and \ref{analysis} detail the rest-frame ultraviolet (UV) spectra collected from the literature and describe our data analysis approach. Section \ref{simulations} outlines our microlensing simulations and discusses the  application of Bayesian methods to infer emission region sizes. \mbox{In Section \ref{results}}, we interpret our findings within a Keplerian framework. Finally, Section \ref{conclusions} summarizes our key findings and presents our main conclusions.

\section{Data and observations}\label{datades}
We have assembled a dataset of rest-frame UV spectra of gravitationally lensed quasars, specifically selecting those exhibiting microlensing effects in the wings of the high-ionization lines Si IV $\lambda$1397 and/or C IV $\lambda$1549. 
Our criteria also required the quasars to have been observed in at least two different epochs and to provide a sufficient signal-to-noise ratio within the wavelength range of these lines. Our dataset includes one quadruply lensed quasar at redshift \mbox{$z = 1.734$}, \mbox{SDSS J1004+4112}, and four doubly lensed quasars, \mbox{SDSS J1001+5027}, \mbox{HE 1104$-$18054}, \mbox{SDSS J1206+4332}, and \mbox{SDSS J1339+1310}, at redshifts $z=1.838$, $2.319$, $1.789$, and $2.243$, respectively. The fully reduced data were obtained from the literature and provided observations spanning multiple epochs for each quasar in our dataset. Specifically, our dataset includes six epochs for SDSS J1001+5027 from 2003 until 2016, 21 epochs for \mbox{SDSS J1004+4112} from 2003 until 2018, seven epochs for HE 1104$-$1805 from 1993 until 2016, two epochs for SDSS J1206+4332 in 2004 and 2016, and eight epochs for \mbox{SDSS J1339+1310} spanning from 2007 until 2017. Information about the observations and references are detailed in Table \ref{AppA}1. In the case of SDSS J1004+4112, we excluded images C and D from our analysis, as their substantial time delays relative to image B ($\sim2$ and $\sim4.5$ years, respectively; see \citealt{Munoz2022}) can confound the interpretation of microlensing. In particular, the intrinsic variability in combination with the long time delays between these image pairs may mimic microlensing effects, leading to erroneous results. It is unlikely that the relatively short time delays between the image pair AB in SDSS J1004+4112 ($\Delta t_{AB} = 44.0\pm0.2$ days, as reported by \citealt{Munoz2022}) and SDSS J1339+1310 ($\Delta t_{AB} = 48\pm2$ days, according to \citealt{Shalyapin2021}) emulate microlensing effects. Care should be exercised when interpreting the magnitude differences observed in the cases of SDSS J1001+5027 and SDSS J1206+4332, as they may arise partially from intrinsic variability compounded by the time delay between the images ($\Delta t = 117.1_{-3.7}^{+7.1}$ and $\Delta t = 111.3_{-3.0}^{+3.0}$ days; as reported in \citealt{Aghamousa2017} and \citealt{Eulaers2013}). However, it is unlikely that intrinsic variability produces the observed asymmetric enhancements in the line profiles (see, e.g., \citealt{Richards2004}).

\section{Data analysis}\label{analysis}
First, we remove the continuum emission around each studied high-ionization line for each image by fitting a straight line to the adjacent continuum regions, carefully avoiding known emission features. Since line widths vary, we use windows of varying widths to estimate the continuum for different lines. The line cores, which are produced by material spread over a wide region (narrow-line region and outer regions of the BLR), are less likely to be affected by microlensing, according to \citet{Fian2018blr}. Therefore, we use them as a reference and establish a baseline for no microlensing by normalizing the emission line cores of images A and B. This normalization involves defining the flux within a narrow interval around the peak of the line and multiplying the spectrum of image B to match the core flux of image A. This procedure also eliminates the effects of macro-magnification and differential extinction (see, e.g., \citealt{Guerras2013}). Next, we build the average of the normalized line profiles for each lensed image of each system under examination, and we fit a spline to the emission line wing that is prone to microlensing. 
For \mbox{SDSS J1001+5027}, we identify a modest microlensing enhancement of image B within the red wing of \mbox{C IV} (see \citealt{Oguri2005}), but we excluded the blue wing from our analysis due to the presence of significant absorption. In the case of \mbox{SDSS J1004+4112}, the microlensing-induced distortions are characterized by a prominent magnification of the blue part of the BELs in image A, and a strong demagnification of the red line part (see \citealt{Hutsemekers2023,Fian2023_1004}). In SDSS J1206+4332 and SDS J1339+1310, a substantial magnification of image B is observable in the red part of the high-ionization lines (see \citealt{Fian2024_SDSS1339,Fian2021blr,Goicoechea2016}), while in HE 1104$-$1805, modest microlensing was detected in the red wings of Si IV and C IV in image A (\citealt{Fian2021blr}). Additionally, in image B of \mbox{SDSS J1206+4332}, magnification is also present in the blue wing of C IV. 

To estimate the differential microlensing between two quasar images at different velocities, we first ascertain the flux ratio between the splines in bins of width 500 km s$^{-1}$ at varying distances from the line core, and subsequently convert this ratio into magnitudes. The BELs in lensed quasars typically exhibit velocities ranging from a few thousand to ten thousand km s$^{-1}$. The exact range of velocities depends on the specific quasar and its characteristics, such as the mass of the central supermassive black hole (SMBH), and the radial extent of the BLR and its inclination with respect to the observer. In this study, we opt to assess the chromatic microlensing by employing velocity intervals of 500 \mbox{km s$^{-1}$}, ranging from 3000 to 8000 km s$^{-1}$, which results in a total of 11 distinct measurements. 
\section{Microlensing simulations}\label{simulations}
\subsection{Magnification Maps}
To model the microlensing phenomenon of spatially extended sources, we employed a novel algorithm, called the Fast Multipole Method -- Inverse Polygon Mapping (FMM--IPM\footnote{https://gloton.ugr.es/microlensing/}), as described by \citet{Jimenez2022}. This technique combines the FMM algorithm of \citet{Greengard1987} for ray deflection calculations and the IPM algorithm of \citet{Mediavilla2006,Mediavilla2011} for magnification map calculations. Our simulations for each quasar image are based on $3000\times3000$ pixel$^2$ maps, covering $100\times100$ Einstein radii$^2$ on the source plane, with a resolution of $0.30$ to $0.40$ light-days per pixel (depending on the Einstein radius, $R_E$, of the quasar). The magnification map for each image was characterized by the fraction of mass in stars $\alpha$, the local shear $\gamma$, and the local convergence $\kappa$, with the latter being proportional to the surface mass density. The local convergence can be split into two components: $\kappa = \kappa_c + \kappa_\star$, where $\kappa_c$ represents the convergence due to continuously distributed matter, such as dark matter, and $\kappa_\star$ represents the convergence due to stellar-mass point lenses, such as microlens stars in the galaxy. The value of the stellar fraction, $\alpha$, is a measure of the relative contribution of stars to the total mass in the lens galaxy. We present the values of $\kappa$ and $\gamma$, as well as $\alpha$, at the position of each image for each lensed quasar system in Table \ref{macromodel}2. To create the convergence listed in \mbox{Table \ref{macromodel}2}, we randomly distributed stars of a mass of $M = 0.3 M_\odot$ across the lens plane.

\subsection{Bayesian Source Size Estimation}\label{size_estimate}
In our study, we use circular Gaussian profiles, $I(R)\propto \exp(-{R}^{2}/2 r_s^{2})$, to model the luminosity of the emitting regions and simulate the structure of the unresolved quasar. To estimate the magnifications experienced by a finite source of size $r_s$, we convolve the magnification maps with 2-D Gaussian profiles with sigma of $r_s$. It is widely accepted that the specific shape of the source's emission profile is not significant for microlensing studies, as the half-light radius controls the results more than the detailed intensity profile\footnote{Nevertheless, Appendix \ref{AppB} explores the kinematic implications of considering an inclined disk within the framework of Keplerian rotation.} (\citealt{Mortonson2005,Munoz2016}; but see \citealt{Fian2023blr}). For Gaussian profiles, the characteristic size $r_s$ relates to the half-light radius as $R_{1/2} = 1.18\, r_s$. Since lengths are measured in Einstein radii, we can rescale all estimated sizes for a different mean stellar mass by applying the relation $r_s \propto \sqrt{M}$. Using the differential microlensing estimates in the wings of the high-ionization lines between lensed images, we can infer the size of the emission region and probe the kinematics of the inner BLR of the quasar. We calculate the probability of reproducing the observed microlensing by randomly placing the Gaussian source on our microlensing magnification maps and estimate the corresponding size following the procedures outlined in \citet{Guerras2013}, as well as in \citet{Fian2018blr,Fian2021blr}.

\section{Results and discussion}\label{results}
\subsection{Velocity-Resolved Responses to Microlensing}

\begin{figure*}
\centering
\includegraphics[width=0.72\textwidth]{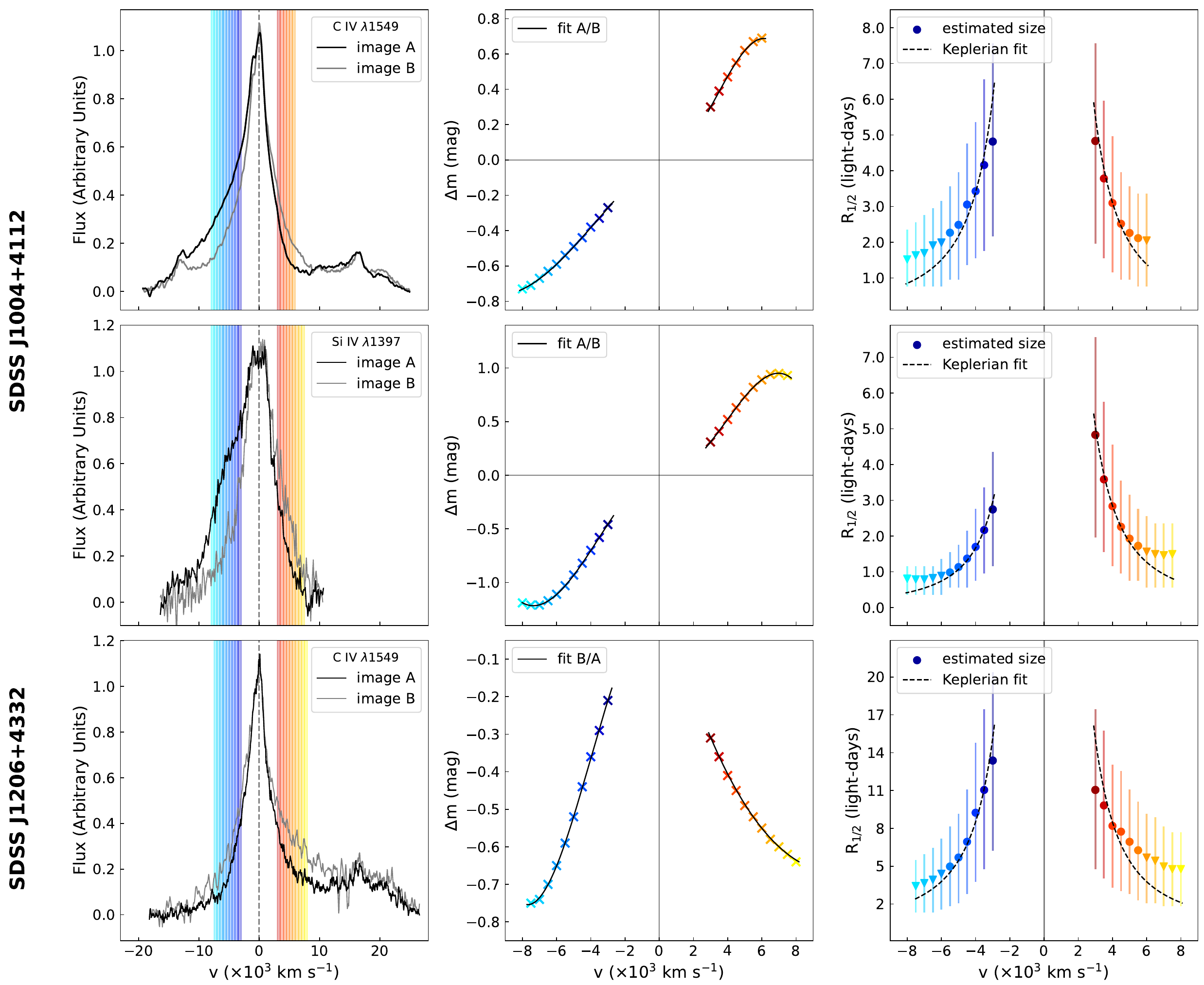}
\caption{Kinematic responses to microlensing for SDSS J1004+4112 and SDSS J1206+4332. The left panels show the average C IV and \mbox{Si IV} line profiles for images A (black) and B (gray). The middle panels illustrate the magnitude differences between the images in the line wings. The rightmost panels display the estimated sizes of different kinematic regions as a function of velocity, along with a Keplerian fit (dashed black line). Upper size limits are shown as triangles and are excluded from the Keplerian fit.}
\label{bothwings}
\vspace*{3mm}
\includegraphics[width=0.72\textwidth]{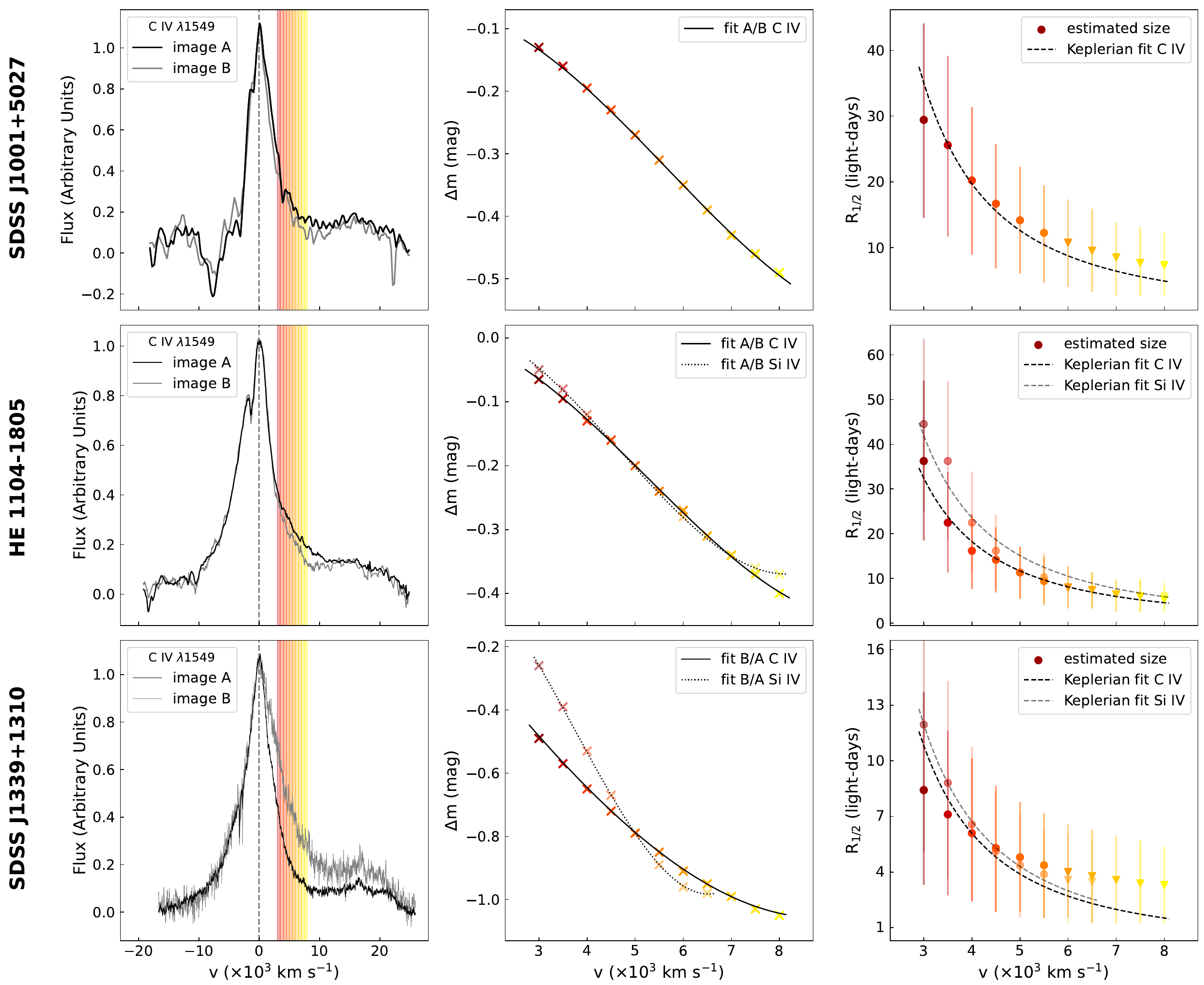}
\caption{Same as Figure \ref{bothwings}, but focusing on the red line wings of the systems SDSS J1001+5027, HE 1104$-$1805, and SDSS J1339+1310. For HE 1104$-$1805 and SDSS J1339+1310, the middle and rightmost panels show the results for both the C IV and Si IV lines.}
\label{redwings}
\end{figure*}
Our study examines the microlensing magnification res\-pon\-se across velocity bins in the wings of the high-ionization lines Si IV and C IV. These regions are relatively free from other emission lines, providing a clear and uncontaminated signal. Using these velocity-resolved microlensing measurements, we can explore the kinematics at various distances from the central SMBH, spanning from tens\footnote{An exception is SDSS J1004+4112, which has a remarkably small BLR as detailed in \citet{Hutsemekers2023} and \citet{Fian2023_1004}.} of light-days down to the innermost region of the quasar of approximately half a light-day. Figures \ref{bothwings} and \ref{redwings} display the average line profiles of images A and B for the five systems in our sample together with the magnitude differences between the images in the microlensing-prone wings as a function of velocity. The first noteworthy finding of our study is the observed monotonic and smooth increase in microlensing \mbox{(de-)}magnification with velocity, which supports the hypothesis that high-velocity emitters originate from smaller regions. We estimate the sizes of the emitting regions of different kinematic regions as outlined in \mbox{Section \ref{size_estimate}}, and present the spatially resolved velocity curves in the rightmost panels of Figures \ref{bothwings} and \ref{redwings}. 

Another significant finding is that when both the red and blue wings of an emission line, or different emission lines from the same object, are analyzed, there is a good coincidence between the velocity curves derived from each wing or emission line. This result strongly validates the consistency of our method and the kinematic conclusions that can be drawn from the velocity curves.

The third notable result is the differing magnification of the blue and red wings across all sources. In some cases, one wing is magnified while the other is de-magnified. This indicates that although both wings originate from regions of comparable size, they are not co-spatial.

Yet another outstanding result of our analysis is the striking kinematic consistency between the velocity curves and the Keplerian model for the systems \mbox{SDSS J1001+5027} and HE 1104$-$1805. The agreement is less pronounced when examining the velocity curves for SDSS J1004+4112, \mbox{SDSS J1206+4332}, and SDSS J1339+1310. In these systems, we observe that the curves tend to be, to varying extents, flatter than predicted by the theoretical model, particularly at the high-velocity end. The deviation from the Keplerian model is most apparent for small sizes, likely due to the flatness of the probability density functions (PDFs) at these scales, which causes a size overestimation. This issue becomes particularly pronounced under conditions of high microlensing magnification. Therefore, for small size determinations, it is more accurate to establish only an upper size limit\footnote{This does not imply a rejection of Keplerian rotation. Instead, it indicates that the PDF of the size distribution is relatively flat, preventing us from providing a precise estimate.}. We note that in the high-velocity wing redward of \mbox{C IV}, a shelf-like feature at $\sim\lambda1610$ is present, which appears with different intensities in different epochs. The origin of this feature is currently uncertain, and possible explanations include an extreme C IV red wing or He II blue wing, or an as-yet-unidentified species (see, e.g., \citealt{Fine2010}). We want to point out that this feature could potentially impact the accuracy of our microlensing measurements and hence affect the estimated sizes in these cases. However, in the blue wing of C IV and both wings of Si IV, we observe similar deviations from the Keplerian fit at high velocities. Notably, these regions do not exhibit a red shelf. Additionally, a non-self-similar cutoff at the inner edge of the BLR (see \citealt{Czerny2011}) might alter the slope at the high-velocity end. However, the discrepancies between the observed and theoretical curves fall within the expected range of uncertainties, indicating that they are still consistent with the Keplerian rotation model. This finding is crucial as it not only links the innermost part of the BLR with the accretion disk but also demonstrates that this region follows Keplerian rotation across a broad range of radial distances. An alternative scenario where a different rotation curve combined with radial motions produces the observed behavior is highly unlikely.\\

To effectively use microlensing for inferring the location of emitters, it is important to note that microlensing primarily measures the spatial extent of the emitting regions of the lensed quasar, rather than directly providing the distances from these regions to the quasar's center. It is crucial to consider the expected scaling between this effective microlensing size and the velocity bin in which it is measured. The most general case is complex, but we have examined a simple disk model with Keplerian rotation in Appendix \ref{AppB}. Surprisingly, we found that under certain conditions, the scaling derived closely aligns with the typical radius versus velocity relationship expected for a Keplerian rotation curve. Therefore, given the reasonable circumstances described in Appendix \ref{AppB}, it is justified to apply the standard Keplerian equation that relates velocity and radius to the mass of the central SMBH, though numerical results should be interpreted with caution.

\subsection{Constraints on SMBH masses from Velocity Curves}
Under the Keplerian hypothesis, fitting the velocity curves allows us to estimate the mass of the central SMBH, $M_{BH}$, multiplied by a factor that depends on the unknown inclination angle of the disk relative to our line of sight, $i$:
\begin{equation}
M_{BH} \sin^2i = \frac{R\ v^2 }{G},
\label{kepler}
\end{equation}

\noindent where $v$ is the observed velocity of the gas orbiting the SMBH, $R$ is the distance of the emitting region from the SMBH, and $G$ is the gravitational constant. The inclination angle affects the observed velocity, with a face-on disk resulting in the lowest observed velocity, and an edge-on disk resulting in the largest observed velocity. In Table \ref{SMBHkepler} we list the inferred masses of the central SMBHs multiplied by $sin^2 i$. By comparing these values with the SMBH mass estimates for each object with available mass measurements from recent literature, we derive an estimate for the mean BLR inclination angle of our sample. We used $\log_{10}(M/M_\odot) = 7.0_{-0.5}^{+0.3}$ for SDSS J1004+4112 (\citealt{Hutsemekers2023}), $\log_{10}(M/M_\odot) = 8.87 \pm 0.70$ for \mbox{HE 1104$-$1805} (\citealt{Melo2023}), $\log_{10}(M/M_\odot) = 8.62 \pm 0.25$ for \mbox{SDSS J1206+4332} (\citealt{Birrer2019}),  and $\log_{10}(M/M_\odot) = 8.6 \pm 0.4$ for \mbox{SDSS J1339+1310} (\citealt{Shalyapin2021}). From these comparisons, we infer that the average inclination angle of the quasar BLR relative to the observer's line of sight is approximately 17$^\circ$. This estimation aligns well with the BLR inclination angles determined for a sample of Seyfert 1 galaxies, as reported by \citet{Williams2018}.

\begin{table}
\tabcolsep=0.335cm
\renewcommand{\arraystretch}{1.2}
\caption{SMBH Mass Estimates multiplied by the Disk Inclination.}\vspace*{-6mm}
\begin{center}
\begin{tabular}{cccc} \hline \hline
Object & Line & Wing & $M_{BH}\, sin^2 i$ ($M_\odot$) \\ 
(1) & (2) & (3) & (4)  \\ \hline 
SDSS J1001+5027 & C IV & red & $6.1_{-1.2}^{+1.5} \times\, 10^7$ \\ \hline
\multirow{4}{*}{SDSS J1004+4112} & \multirow{2}{*}{C IV} & blue & $1.1_{-0.4}^{+0.7} \times\, 10^7$ \\ 
& & red & $1.0_{-0.4}^{+0.7} \times\, 10^7$ \\ 
& \multirow{2}{*}{Si IV} & blue & $0.5_{-0.3}^{+0.6} \times\, 10^7$ \\
& & red & $0.9_{-0.4}^{+0.7} \times\, 10^7$ \\ \hline 
\multirow{2}{*}{HE 1104$-$1805} & C IV & red & $5.7_{-1.2}^{+1.5} \times\, 10^7$ \\ 
& Si IV & red & $7.4_{-1.4}^{+1.7} \times\, 10^7$ \\  \hline 
\multirow{2}{*}{SDSS J1206+4332} & \multirow{2}{*}{C IV} & blue & $2.7_{-0.8}^{+1.1} \times\, 10^7$ \\ 
& & red & $2.6_{-0.8}^{+1.1} \times\, 10^7$ \\ \hline 
\multirow{2}{*}{SDSS J1339+1310} & C IV & red & $1.9_{-0.6}^{+0.9} \times\, 10^7$ \\ 
& Si IV & red & $2.1_{-0.7}^{+1.0} \times\, 10^7$ \\ \hline 
\end{tabular}
\end{center}\vspace*{-1mm}
{\small NOTES. --- Col. (1): Lensed quasar. Cols. (2)--(3): Emission line and wing used for estimating the spectral responses to microlensing. Col. (4): SMBH mass multiplied by the disk inclination (Eq. \ref{kepler}).}
\label{SMBHkepler}    
\end{table}

\subsection{Composite Velocity Curve}
\begin{figure}
\centering 
\includegraphics[width=0.47\textwidth]{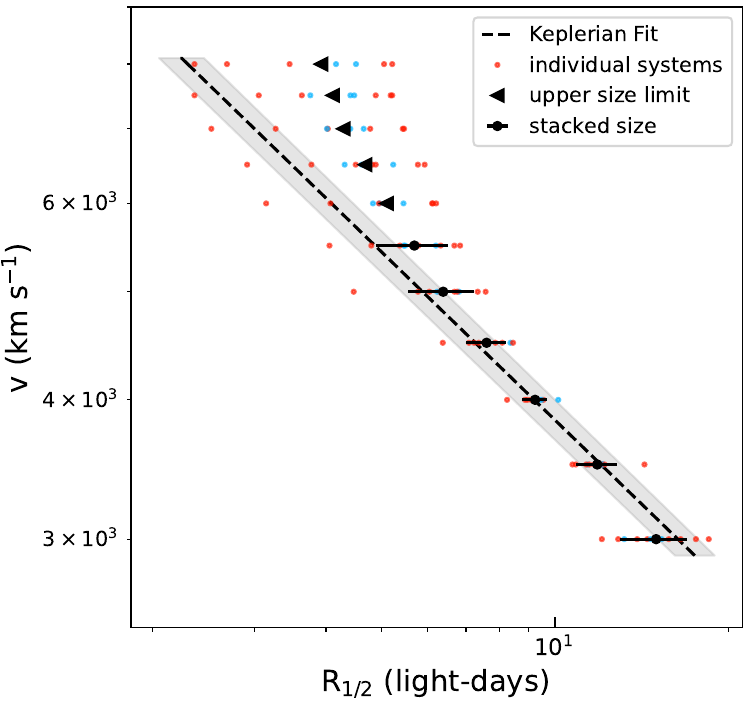}
\caption{Composite velocity-size curve (black data points), along with a Keplerian fit (dashed black line). The gray-shaded area represents the one-sigma uncertainty interval of the Keplerian fit. The velocity curves for individual systems and/or emission lines are represented by blue and red data points, which correspond to the blue and red wings, respectively. Upper size limits are shown as triangles and are excluded from the Keplerian fit.}
\label{stack}
\end{figure}

Figure \ref{stack} presents the composite velocity curve derived from stacking the individual velocity curves of all systems, with the emission region sizes scaled to a mean black hole mass obtained from the 11 individual measurements ($M_{BH} sin^2 i = 2.9\times 10^7 M_\odot$; see Table \ref{SMBHkepler}). The observed kinematics predominantly indicate near-Keplerian motion within the velocity range of 3000 to 5500 km s$^{-1}$. As discussed for the individual velocity curves, deviations occur at small distances/high velocities (6000 to 8000 km s$^{-1}$), where the data follow a flatter slope compared to the predicted Keplerian curve. We note that the apparent large scatter for small emitting region sizes is a result of the logarithmic scale used. Despite the aforementioned deviations, the overall good fit provides strong evidence for the existence of an inner BLR component with characteristics consistent with a disk-like geometry. There is no significant indication of an additional outflow component.

\section{Conclusions}\label{conclusions}
We have obtained velocity-resolved microlensing magnification curves for the wings of the C IV and Si IV BELs in five gravitationally lensed quasars: SDSS J1001+5027, \mbox{SDSS J1004+4112}, \mbox{HE 1104$-$1805}, SDSS J1206+4332, and SDSS J1339+1310. The wavelength regions sampled in the wings are thought to correspond to the innermost parts of the BLR. By utilizing the strength of microlensing to estimate the size of the emitting regions for each velocity bin, we have derived velocity-size curves. We observed a monotonic and smooth increase in microlensing magnification (indicating a decrease in size) with velocity, confirming the hypothesis that high-velocity emitters originate from smaller regions. Significant kinematic consistency is found when these velocity-size relationships are derived from either the red or blue wings of the same emission line, or from different emission lines of the same object. Both individual and composite velocity curves align closely with the Keplerian model of an inclined disk. This remarkable result associates the innermost part of the BLR directly with the accretion disk. Consequently, we demonstrate for the first time that the innermost region of the BLR in quasars follows Keplerian rotation across a broad range of radial distances, from $\sim$5 to 20 light-days. Alternative explanations (e.g., involving radial motions) would require extremely precise fine-tuning to replicate the observed velocity curves. Finally, we would like to emphasize the unprecedented spatial resolution achieved -- below one light-day -- made possible by the extraordinary sensitivity of gravitational microlensing to size.

\section*{Acknowledgments}
This research was supported by the grants PID2020-118687GB-C31, PID2020-118687GB-C32, and PID2020-118687GB-C33, financed by the Spanish Ministerio de Ciencia e Innovación. J.J.V. is also financed by the project FQM-108, financed by Junta de Andalucía. D.C. and S.K. are financially supported by the DFG grant HA3555-14/1 to Tel Aviv University and University of Haifa, and by the Israeli Science Foundation grant no. 2398/19 and 1650/23.

\bibliography{sample631}{}
\bibliographystyle{aasjournal}

\appendix \vspace*{-0.5cm}
\renewcommand{\thetable}{A\arabic{table}}
\setcounter{table}{0}
\section{Supplementary material}\label{AppA}
\begin{table*}[h]
\begin{center}
\tabcolsep=0.53cm
\renewcommand{\arraystretch}{0.86}
\caption{Spectroscopic Data.}
\begin{tabular}{cccccc} \hline \hline 
Object & Date & Line & Image & Facility & Reference\\ \hline 
\multirow{6}{*}{SDSS J1001+5027} & 20-11-2003 & C IV & A, B & ARC 3.5m & \citealt{Oguri2005}\\
& 28-01-2013 & C IV & A, B & MMT 6.5m & E. Falco (private comm.) \\
& 26-02-2015 & C IV & A, B & LT 2.0m & \citealt{Gil-Merino2018}\\
& 02-12-2015 & C IV & A, B & LT 2.0m & \citealt{Gil-Merino2018}\\
& 05-04-2016 & C IV & A, B & LT 2.0m & \citealt{Gil-Merino2018}\\
& 06-12-2016 & C IV & A, B & LT 2.0m & \citealt{Gil-Merino2018}\\ \hline
\multirow{20}{*}{SDSS J1004+4112} & 03-05-2003 & C IV & A & ARC 3.5m & \citealt{Richards2004}\\ 
& 03-02-2003 & C IV & B &  SDSS 2.5m & \citealt{Richards2004} \\
& 31-05-2003 & Si IV, C IV & A, B &  Keck I 10m & \citealt{Richards2004} \\
& 21-11-2003 & C IV & A, B & ARC 3.5m & \citealt{Richards2004} \\
& 30-11-2003 & C IV & A, B & ARC 3.5m & \citealt{Richards2004} \\ 
& 01-12-2003 & C IV & A, B & ARC 3.5m & \citealt{Richards2004} \\ 
& 22-12-2003 & C IV & A, B & ARC 3.5m & \citealt{Richards2004} \\ 
& 19-01-2004 & C IV & A, B & WHT 4.2m & \citealt{Gomez2006}\\ 
& 26-03-2004 & C IV & A, B & ARC 3.5m & \citealt{Richards2004b}\\ 
& 10-04-2004 & C IV & A, B & ARC 3.5m & \citealt{Hutsemekers2023}\\ 
& 26-04-2004 & C IV & A, B & ARC 3.5m & \citealt{Hutsemekers2023}\\ 
& 13-05-2004 & C IV & A, B & ARC 3.5m & \citealt{Hutsemekers2023}\\ 
& 28-05-2004 & C IV & A, B & ARC 3.5m & \citealt{Hutsemekers2023}\\ 
& 08-12-2004 & C IV & A, B & ARC 3.5m & \citealt{Hutsemekers2023}\\
& 17-12-2004 & C IV & A, B & ARC 3.5m & \citealt{Hutsemekers2023}\\
& 01-05-2006 & C IV & A, B & ARC 3.5m & \citealt{Hutsemekers2023}\\
& 16-05-2007 & C IV & A, B & SAO RAS 6m & \citealt{Popovic2020} \\
& 12-01-2008 & Si IV, C IV & A, B & MMT 6.5m & \citealt{Motta2012}\\
& 27-10-2008 & C IV & A, B & SAO RAS 6m & \citealt{Popovic2020}\\
& 11-03-2016 & C IV & A, B & WHT 4.2m & \citealt{Fian2021blr}\\
& 07-02-2018 & C IV & A, B & SAO RAS 6m & \citealt{Popovic2020}\\ \hline
\multirow{7}{*}{HE 1104$-$1805} & 11-05-1993 & Si IV, C IV & A, B & NTT 3.6m & \citealt{Wisotzki1993} \\
 & 18-05-1993 & Si IV, C IV & A, B & AAT 3.9m & \citealt{Smette1995} \\
 & 29-11-1994 & Si IV, C IV & A, B & ESO 3.6m & \citealt{Wisotzki1995} \\
 & 27-02-2003 & Si IV, C IV & A, B & WHT 4.2m & \citealt{Gomez-Alvarez2004} \\
 & 11-01-2008 & Si IV, C IV & A, B & MMT 6.5m & \citealt{Motta2012} \\
 & 07-04-2008 & Si IV, C IV & A, B & VLT 8.2m & \citealt{Motta2012} \\
 & 12-03-2016 & Si IV, C IV & A, B & WHT 4.2m & \citealt{Fian2021blr} \\ \hline 
\multirow{2}{*}{SDSS J1206+4332} & 21-06-2004 & Si IV, C IV & A, B & ARC 3.5m & \citealt{Oguri2005} \\
 & 12-03-2016 & C IV & A, B & WHT 4.2m & \citealt{Fian2021blr} \\ \hline 
\multirow{8}{*}{SDSS J1339+1310} & 13-05-2007 & C IV & A, B & UH88 2.2m & \citealt{Inada2009} \\
& 12-03-2012 & Si IV, C IV & A & SDSS 2.5m & \citealt{Dawson2013}\\
& 29-03-2012 & Si IV, C IV & B & SDSS 2.5m & \citealt{Dawson2013}\\
& 13-04-2013 & C IV & A, B & GTC 10.4m & \citealt{Shalyapin2014}\\
& 27-03-2014 & C IV & A, B & GTC 10.4m & \citealt{Goicoechea2016} \\
& 20-05-2014 & Si IV, C IV & A, B & GTC 10.4m & \citealt{Goicoechea2016} \\
& 18-02-2016 & C IV & A, B & HST 2.4m & \citealt{Lusso2018} \\
& 06-04-2017 & Si IV, C IV & A, B & VLT 8.2m & \citealt{Shalyapin2021} \\ \hline 
\end{tabular}
\end{center}
\small \vspace*{-0.17cm} NOTES. --- AAT: Anglo-Australian Telescope; 
ARC: Astrophysical Research Consortium; 
ESO: European Southern Observatory;
GTC: Gran Telescopio CANARIAS;
HST: Hubble Space Telescope; 
LT: Liverpool Telescope; 
MMT: Multiple Mirror Telescope; 
NTT: New Technology Telescope;
SAO RAS: Special Astrophysical Observatory of the Russian Academy of Sciences;
SDSS: Sloan Digital Sky Survey; 
UH88: University of Hawaii 88\arcsec\ Telescope; 
VLT: Very Large Telescope;
WHT: William Herschel Telescope.
\normalsize
\label{data}    
\end{table*}

\begin{table*}[h]
\begin{center}
\tabcolsep=0.59cm
\renewcommand{\arraystretch}{0.91}
\caption{Macro-Model Parameters.} \vspace*{-0.3cm}
\begin{tabular}{cccccccc} \hline \hline 
Object & $R_E$ ($\times10^{16}$ cm) & $\kappa_A$ & $\kappa_B$ & $\gamma_A$ & $\gamma_B$ & $\alpha_A$ ($\%$) & $\alpha_B$ ($\%$) \\ 
(1) & (2) & (3) & (4) & (5) & (6) & (7) & (8) \\ \hline 
SDSS J1001+5027$^{\, (a)}$ & $3.08$ & $0.35$ & $0.74$ & $0.28$ & $0.72$ & $21$ & $21$ \\
SDSS J1004+4112$^{\, (b)}$ & $2.35$ & $0.73$ & $0.65$ & $0.33$ & $0.23$ & $5$ & $5$ \\ 
HE 1104$-$1805$^{\, (c)}$ & $2.36$ & $0.64$ & $0.33$ & $0.52$ & $0.21$ & $21$ & $21$ \\
SDSS J1206+4332$^{\, (d)}$ & $3.11$ & $0.43$ & $0.63$ & $0.41$ & $0.72$ & $21$ & $21$ \\  
SDSS J1339+1310$^{\, (e)}$ & $2.97$ & $0.40$ & $0.63$ & $0.49$ & $0.90$ & $28$ & $52$ \\ \hline \vspace*{-3mm} 
\end{tabular}
\end{center}
\small \vspace*{-0.25cm} NOTES. --- Col. (1): Gravitationally lensed quasar. Col. (2): Einstein radius assuming $M = 0.3 M_\odot$ stellar mass objects, as reported in \citet{Mosquera2011}. Cols. (3)--(6): Convergence and shear of images A and B obtained from the following sources: \citet{Mediavilla2009} for (a), (c) and (d), \citealt{Fores2022} for (b), and \citet{Shalyapin2021} for (e). Cols. (7)--(8): Fraction of mass in stars for images A and B obtained from the following sources: a standard estimate near the Einstein radius of the lenses ($\sim 1-2$ effective radii) for (a), (c) and (d) from \citet{Jimenez2015}, \citet{Fores2022} for (b), and \citet{Shalyapin2021} for (e).\normalsize
\label{macromodel}    
\end{table*}

\vspace*{-0.17cm}\section{Size-velocity curve for an inclined rotating disk} \label{AppB}
We explore a simple model of an inclined rotating disk with a Keplerian rotation curve defined by $v(R)=\sqrt{M_{BH}G/R}$. The observed (radial) velocity field for such a disk, with an inclination $i$ with respect to the plane of the sky and with $x$ lying along the major axis of the observed disk, can be described by the following equation:
\begin{equation}
    \label{velfield}
        v_r(x,y)=\frac{\sqrt{GM_{BH}}x\sin{i}}{R'^{3/2}} 
    \end{equation}
where $R'=\sqrt{x^2+y^2/\cos^2{i}}$. We can select the emission region corresponding to a specific velocity bin centered at $v_r$ with a width of $\Delta v_r$ by selecting the (implicitly defined) region between the curves corresponding to the velocities at the edges of the velocity bin, $v_r\pm\Delta v_r/2$. The {\em effective} area of this region can thus be calculated by integrating the surface brightness of the disk, described by its (normalized) radial profile $I(R)$, over that region, denoted as 
$\Sigma$:
\begin{equation}
        \label{area}
        A(v_r\pm\Delta v_r/2)=\int_\Sigma I(R')dS
\end{equation}
Typically, the disk does not extend to $R=0$ in the inner region or to $R=\infty$ in the outer region. Therefore, the region $\Sigma$ is further constrained by the (projected) ellipses that correspond to an annulus with an inner radius $R_{in}$ and outer radius $R_{out}$. The integral described can be calculated numerically. Finally, from this area, we must determine an {\em effective} size corresponding to the half-light radius $R_{1/2}$ of the emitting region, a measurement derived from microlensing. For our purposes, we approximate $R_{1/2}\approx \sqrt{A/2}$. We recognize that a different proportionality constant might be necessary depending on factors such as the specific shape of the region, but this approximation should be adequate for current analyses. Even for such a simple model, there are several free parameters to consider. Fully exploring the entire parameter space is beyond the scope of this work, as our aim is primarily to gain insights into expected behaviors. Therefore, we focus on the simplest scenario -- a uniform disk (ring). However, we will briefly comment on the implications of relaxing this assumption at the end of this section.

We find that in the scenario where $M_{BH}=1.6\times 10^8 M_\odot$, $i=30^\circ$, $R_{in}=1$ light-day and $R_{out}=100$ light-days, the size versus velocity curve follows closely a power law ($R_{1/2} \propto v^{-2.5}$) very similar to the expected Keplerian rotation curve ($R \propto v^{-2}$). Consequently, using the standard Keplerian formula provides a reasonable approximation. 
A few words of caution are warranted here. First, the model presented is an extreme simplification, and the actual morphology and kinematics of the BLR are likely more complex, potentially including an extended region dominated by random motions. We are, to some extent, assuming that microlensing primarily affects the smallest, innermost regions. However, even within this highly simplified framework, numerous parameters can influence the outcomes. For instance, varying radial brightness profiles (including a characteristic scale) can alter the slope of the size versus velocity relationship (i.e., steeper brightness profiles tend to yield shallower relations). On the other hand, the intercept of this relation -- which relates to the mass of the black hole -- strongly depends on the precise definition of the {\em effective} size. We have approximated this size as $R_{1/2}\approx \sqrt{A/2}$. Additionally, we must consider the potential for non-Keplerian rotation influencing our results. Consequently, while using a Keplerian rotation equation to simplify the size versus velocity relationship is tempting and somewhat justified, the obtained numerical values must be interpreted with caution. Despite these caveats, the fact that the observed relations closely resemble the simplified model suggests that this description captures some aspects of the actual scenario. Thus, it implies that the inner part of the BLR, which is most sensitive to microlensing, likely includes a substantial contribution from a rotating structure.
\end{document}